\documentclass[sigconf]{acmart}

\makeatletter                   
\def\mdseries@tt{m}             
\makeatother                    
\usepackage[plain]{fancyref}
\usepackage[draft=true]{minted} 
\usepackage{color}

\usepackage{breakurl}           

\renewcommand\footnotetextcopyrightpermission[1]{} 
\usepackage[shortcuts]{extdash}
\usepackage{times}
\usepackage{amsmath}
\usepackage{amsmath,amssymb,amsfonts}
\usepackage{algpseudocode,algorithmicx}
\usepackage[ruled]{algorithm2e} 
\usepackage{algpseudocode,algorithmicx}

\usepackage{graphicx}
\usepackage{textcomp}
\usepackage{xcolor}
\usepackage{comment}
\usepackage{caption}
\usepackage{url}
\usepackage{frame}

\usepackage[export]{adjustbox}

\setcopyright{none}

\newcommand{\name}{{\textit{DistCLUB}}}
\newcommand{\perf}{{8.87x}}
\newcommand{\accr}{14.5\%}

\begin{document}

\title{Fast Distributed Bandits for Online Recommendation Systems}

\author{Kanak Mahadik}
\affiliation{
  \institution{Adobe Research}
  \city{San Jose}
  \country{USA}}
\email{mahadik@adobe.com}

\author{Qingyun Wu}
\affiliation{
  \institution{University of Virginia}
 \city{Charlottesville}
  \country{USA}}
\email{qw2ky@virginia.edu}

\author{Shuai Li}
\affiliation{
  \institution{Cambridge University}
  \city{Cambridge}
  \country{United Kingdom}}
\email{shuaili.sli@gmail.com}

\author{Amit Sabne}
\affiliation{%
  \institution{Google Brain}
  \city{Mountain View}
  \country{USA}}
\email{asabne@google.com}


\begin{abstract}

Contextual bandit algorithms are commonly used in recommender systems, where content popularity can change rapidly. These algorithms continuously learn latent mappings between users and items, based on contexts associated with them both. Recent recommendation algorithms that learn clustering or social structures between users have exhibited higher recommendation accuracy. However, as the number of users and items in the environment increases, the time required to generate recommendations deteriorates significantly. As a result, these cannot be deployed in practice. The state-of-the-art distributed bandit algorithm - DCCB - relies on a peer-to-peer network to share information among distributed workers. However, this approach does not scale well with the increasing number of users. Furthermore, it suffers from slow discovery of clusters, resulting in accuracy degradation. 

To address the above issues, this paper proposes a novel distributed bandit-based algorithm called {\name}. This algorithm lazily creates clusters in a distributed manner, and dramatically reduces the network data sharing requirement, achieving high scalability. Additionally, {\name} finds clusters much faster, achieving better accuracy than the state-of-the-art algorithm. Evaluation over both real-world benchmarks and synthetic data\-sets shows that {\name} is on average {\perf} faster than DCCB, and achieves {\accr} higher normalized prediction performance.

\end{abstract}



\maketitle

\section{Introduction}

Large-scale recommendation systems are deployed in businesses such as retail, video-on-demand, and music streaming. They are fundamental drivers for web services such as YouTube, Netflix, Amazon, and Google News. Recommendation systems provide the user with suggestions on what they should watch next, shop next, or do based on the time of the day. According to Mckinsey, the recommendation feature contributed to 35\% and 23.7\% growth in revenue for Amazon and BestBuy respectively. The report also stated that 75\% of the video-consumption and 60\% views on web services Netflix and YouTube~\cite{businessimpact} respectively are due to recommendations. Hence, developing effective recommendation systems is extremely important for businesses such that appropriate content is suggested to users in a timely manner.

Recommendation systems are deployed in social environments where user preferences and the set of active users change with time. In addition, many web services comprise content that is not only dynamic, but also receives rapid updates. Traditional recommendation systems involve supervised learning algorithms that use collaborative and content-based filtering methods~\cite{he2016fast,garcin2014offline}. The key issue faced by these supervised learning approaches is that they train \emph{offline}. Real applications, however, face a phenomenon called \textit{concept drift}~\cite{tsymbal2004problem,gama2014survey,schlimmer1986beyond}, wherein user preferences and content could change within minutes. Concept drift makes results obtained by supervised learning sub-optimal. The large-scale at which these web services operate make it necessary to have algorithms that learn fast, in an \emph{online} fashion. Bandit-based algorithms can adapt to these updates and therefore are prevalent in online settings~\cite{li2010contextual,gentile2014online}. A \textit{bandit learner} or \textit{agent} picks an action from a set, i.e., an item to recommend to the user, and observes the user feedback to learn about her preferences for the set of items. If the feedback is positive, e.g. the user clicks on the item, the agent receives a reward. The agent is inclined to pick actions that provide higher rewards, thus reinforcing better user-item mappings.

As the number of users and items increases, these bandit-based algorithms face severe challenges in scaling up to satisfy the requirements of these modern applications that cater to millions of users with billions of possible recommendable items. High-quality, speedy, and dynamic recommendations are critical to the success of these enterprises. At scale, the data sizes and compute volumes go well beyond the capacity of a single compute node. Therefore, the ability to compute recommendations in a distributed manner becomes paramount~\cite{hillel2013distributed}.

Unfortunately, most of the bandit-based recommendation algorithms are neither parallel nor distributed~\cite{gentile2014online, li2010contextual, mcinerney2018explore}. Bandit-based algorithms access and edit dynamic user and item information, necessitating data coherence across different cores/nodes. The resultant high synchronization costs limit the scalability of a distributed approach. The state-of-the-art distributed online recommendation algorithm, called Distributed Clustering of Confidence Ball (DCCB)~\cite{korda2016distributed} builds upon the methodology used in its sequential counterpart, named Cluster of Bandits (CLUB)~\cite{gentile2014online}.
CLUB operates by organizing users into different clusters and recommending items to users based on the cluster's cumulative properties. The process of cluster formation is dynamic, and therefore adaptable to changing user preferences. If a distributed algorithm, such as DCCB, had to dynamically create clusters and perform recommendations, it would need all-to-all communications after each interaction. DCCB overcomes this expensive need by a) updating clusters in a lazy, periodic fashion and b) it archives the interactions-generated user feedback for the period in a buffer. After every period, each node shares its buffer with a randomly chosen peer so as to propagate the newly learned user-item mappings. Bandit-based algorithms are resilient to such lazy updates~\cite{korda2016distributed}. 

In spite of the lazy distributed updates, DCCB runs into scalability issues owing to its reliance on peer-to-peer communications transferring large data buffers. To reduce communication penalties, DCCB transfers the data periodically. However, this results in slow discovery of clusters and a reduction in accuracy since the restricted peer-to-peer information exchange introduces bias and imperfections due to partial information visibility.
The communication bottleneck and imperfect information sharing exacerbate in environments with large user counts. Because each machine must handle several users, the buffer space gets constrained, requiring more frequent communication.

In this work, we present {\name} \footnote{Our software is available at \textit{https://bit.ly/distclub}}, a novel distributed algorithm that overcomes these issues. {\name} eliminates the need for large buffer transfers by distributing execution across nodes based on user-level and cluster-level parallelism. It employs efficient reduction primitives, boosting the cluster discovery rate. {\name} adaptively alternates between recommending based on user and cluster information. It leverages heuristics to statistically estimate the volume of interaction data available for each user, and \textit{intelligently} balances highly personalized and cluster-based recommendations. The reduced communication overheads allow {\name} to quickly discover clusters and achieve higher accuracy than DCCB. The paper presents a detailed regret analysis for {\name} which shows that it achieves the same asymptotic regret bound as DCCB and CLUB.

To the best of our knowledge, {\name} is the first large-scale, high-performance distributed system for online recommendations that can scale to 64 nodes (512 cores). {\name} is built using Spark~\cite{zaharia2010spark}, and carefully maps operations on Spark's distributed primitives. We evaluate {\name} on several real-world applications, achieving an average execution speedup of {\perf} while obtaining {\accr} better normalized prediction performance than DCCB.

This paper makes the following contributions:

\begin{enumerate}
    \item The paper presents a novel distributed algorithm named {\name} that provides online recommendations at scale.
    
    \item The paper designs and implements a Spark-based distributed system to demonstrate the effectiveness of \name.
    
    \item The paper experimentally demonstrates that {\name} can scale to millions of user interactions, while operating faster and achieving better accuracy than the state-of-the-art DCCB  approach.
    
\end{enumerate}

The rest of this paper is organized as follows. Section~\ref{sec:algos} describes the state-of-the-art online recommendation algorithm, CLUB and DCCB. Section~\ref{sec:design} present the {\name} algorithm and its system design. Section~\ref{sec:analysis} describes the theoretical regret analysis of {\name}.
Section~\ref{sec:eval} presents experimental results. Section~\ref{sec:related} overviews the related work, while section~\ref{sec:conclusion} concludes the paper.
\section{Bandit-based Online Recommendation Algorithms}
\label{sec:algos}

\begin{figure}[h]
\centering
\hspace{-0.5cm}
\includegraphics[width=0.9\linewidth]{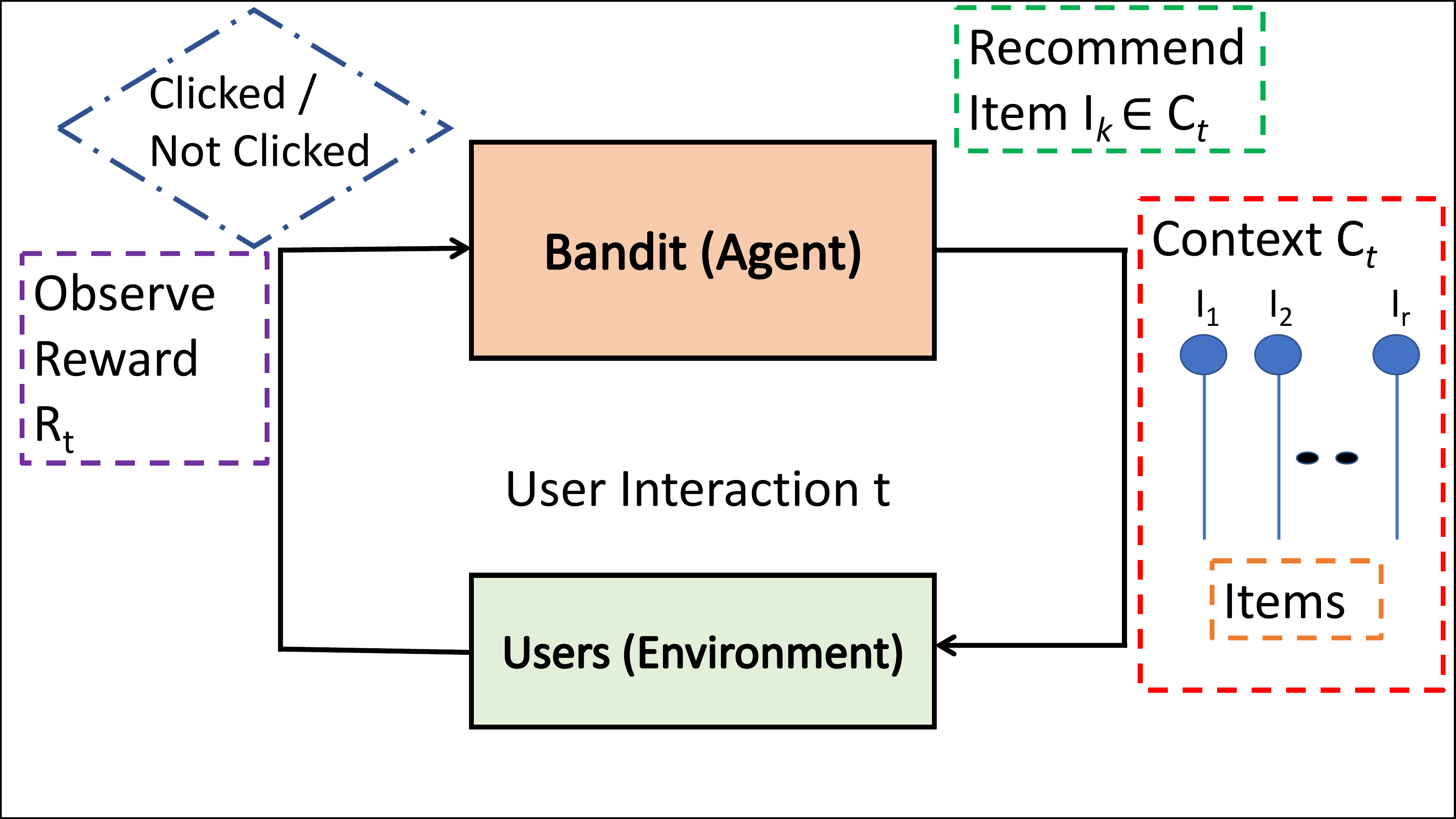}
\caption{Bandit-based Recommendation System}
\label{fig:bandit}
\end{figure}

Figure~\ref{fig:bandit} shows a typical bandit-based recommendation system. For each interaction \textit{t}, the bandit agent picks an action, i.e. an item $I_{k}$ to recommend to the user. Context $C_{t}$ represents user and items features. For example, in a movie recommendation system, user features could be geographical information, historical behaviour while item features could be genre, rating, etc. 
The action of selecting $I_{k}$ is based on context $C_{t}$ for this interaction, and feedback obtained in the previous interactions~\cite{agarwal2014taming}. The problem of sequentially selecting an item among multiple items with each having an unknown reward  distribution is an instance of \textit{multi-armed bandit} problem\footnote{The term comes from a casino setting wherein a player is facing multiple slot machines (\textit{arms}) at once, and needs to repeatedly choose where to insert the next coin. "Bandit" is an American slang for slot machine.}. The user responds to this recommendation by either clicking on it or not, which is considered as the reward $R_{t}$ to the agent . If the user clicks the item the agent receives a non-zero reward for the item. Based on the user feedback, the agent learns about user preferences for the set of items. Note that the agent only observes the reward for the recommended item.  

This section first describes the sequential Cluster of Bandits  (CLUB)~\cite{gentile2014online}, followed by the state-of-the-art distributed online recommendation algorithm named Distributed Clustering Confidence Ball (DCCB)~\cite{korda2016distributed}.

\begin{algorithm}[!htb]
\SetAlgoLined
\fontsize{7.7}{10}\selectfont
\KwIn{$\alpha$ : UCB hyperparameter}
\KwIn{$\delta$ : Network update delay}  
\KwIn{$\gamma$ : cluster-partitioning hyperparameter}
\KwIn{$G(V,E):$ Graph where vertices represent users. Edges represent correlation between users. Initially it is a complete graph. }
\KwIn{$occ$: Array of counters representing interaction occurrences for the users , initialized to 0}
\KwIn{context: table with features for all recommendable items}
\textbf{Initial state:}
 $Mu[users] = I  \qquad \forall users \in [0, numUsers]$\\
 $\vec{bu}[users] = 0  \qquad \forall users \in [0, numUsers]$\\
 
\hrulefill
\SetKwFunction{Function}{UCB}
\Function{$\vec{w}$, $occ$, $context$, $MInv$}\\
{    \For{each item k in context}{
        estimate $\gets$ {k} $\times \vec{w}$\\
        bonus $\gets  \alpha \times \sqrt{k^{\top} \cdot  MInv \cdot k } \times \sqrt{\log(1+occ)}$ \\
        UCB-VAL[k] = estimate + bonus \\
        }

    \Return   argmax(UCB-VAL)
}
\hrulefill
\\
\SetKwFunction{Function}{updateNetwork}

\Function{ }\\
{
\For{each user pair ($u_1, u_2$)}{
    \If{exists E[$u_1$, $u_2$]}{
        $\vec{v_{1}} \gets  {Mu[u_1]}^{-1}\times \vec{bu}[u_1] $ \\
        $\vec{v_{2}} \gets  {Mu[u_2]}^{-1}\times \vec{bu}[u_2] $ \\
         \If{$|\vec{v_{1}} - \vec{v_{2}}|\geq \gamma \times$ {getThreshold($\vec{v_{1}}$, $\vec{v_{2}}$)}}{
            remove $E[u_1,u_2]$ \\
        }
    }
}
}
\hrulefill
\For{each user interaction i}{
    {userID} $\gets$ getUserIdFromInteraction(i) \\
    {clusterID} $\gets$ clusters[userID]\\
    {Mc} $\gets$ I, $\vec{bc}$ $\gets$ 0\\
    \For{each user n $\in$ {clusterID}\\}{ 
        {Mc} += {{Mu[n]}}\\
        $\vec{bc}$ += $\vec{bu}[n]$\\
    }
     $\vec{w} \gets {Mc}^{-1} \times \vec{bc}$\\
    
    choice $\gets$ {UCB}($\vec{w}$, occ[userID], getContext(i), ${Mc}^{-1}$) \\
    reward $\gets$ getReward(choice)\\
    \textcolor{blue}{// standard linear bandit algorithm}\\
    {Mu}[userID] += {context[choice]} $\cdot$ ${context[choice]}^\top$\\
    $\vec{bu}$[userID] += reward $\times$ context[choice] \\
    \If{i $\%$ $\delta$ == 0\\}{
        updateNetwork()\\
        clusters $\gets$ recomputeClusters(G)\\

    }
    {occ[userID]}  $\gets$ {occ[userID]} + 1 \\

}
\caption{Cluster of Bandit (CLUB) - UCB-based clustering; a sequential approach to online recommendations}
\label{algo:club}

\end{algorithm}

\subsection{Cluster of Bandits (CLUB)}

Listing~\ref{algo:club} shows the CLUB \cite{gentile2014online} algorithm. It sequentially clusters users via linear contextual bandits. 

CLUB begins with a strongly connected graph. Through the incoming user interactions, it partitions the users into $m$ clusters such that $m \ll n$.


Let $G(V,E)$ be the undirected graph consisting of vertices $V$ and edges $E \subseteq V \times V$. Each vertex $\{u_{1},u_{2},...,u_{n}\} \in V$ represents a user, and an edge between two vertices indicates that the users have similar preferences. 

Features of items in the recommendation system are encoded using a $d$-dimensional vector. The bandit agent maintains a $d$-dimensional user-vector for each user which encapsulates the user's preference for these features of the items. In other words, these item vectors, denoted as $\{\vec{v_{1}},\vec{v_{2}},..,\vec{v_{n}}\}$, are the agent's estimation of the user's preferences. In the linear contextual bandit setting \cite{abbasi2011improved}, $\vec{v_{i}}$ can be obtained by multiplying the inverse of correlation matrix $M_{i}$ and the bias vector $b_{i}$~\cite{gentile2014online}. Cluster vectors are calculated by aggregating user-vectors in the same cluster. For $m$ clusters, we denote the cluster vectors as  $\{\vec{w_{1}},\vec{w_{2}},..,\vec{w_{m}}\}$.

A user interaction entails a bandit learner recommending an item to the user, and observing their reaction to it, obtained as a \textit{reward}. During each interaction, the learner recommends an item from an item-list, i.e., context, as illustrated in Figure~\ref{fig:bandit}. 

Upper Confidence Bound (UCB) bandit algorithm~\cite{auer2002finite,auer2002using,lai1987adaptive} estimates an upper confidence bound for each arm as the sum of the sample mean reward  and the confidence interval (exploration bonus). It then picks the arm with the highest calculated upper confidence bound on the reward for a user on each interaction. Note that UCB associates higher confidence to users that were previously interacted with, i.e. confidence intervals for an arm decrease after its true reward is observed. Importantly, estimated cluster vectors are used to make recommendations and not user-vectors. After observing the obtained reward, CLUB updates the user-vector according to the contextual choice using a standard linear-bandit algorithm~\cite{chu2011contextual}.


Clusters are recomputed according to the updated user vectors. The key principle exploited by the algorithm for clustering is that similar users have similar ground truth vectors. If the difference between two vectors is greater than a threshold,  the corresponding users are considered dissimilar and are disconnected from each other.  In practice, recomputing clusters after each interaction becomes expensive, and hence is typically done once in a few interactions.

\begin{algorithm}[!htb]
\fontsize{7.7}{10.3}\selectfont
\SetAlgoLined
\KwIn{$G(V,E):$ User Graph, All users in the same cluster initially.}

\KwIn{{MBuff}, {bBuff} - Buffers of length bufferSize, Mw, $\vec{bw}$ - current copies, $Mw_{local}$, $\vec{bw}_{local}$ - local working copies }

\textbf{Initial state:} Mw{[users]}= I $\qquad$ $\forall$ users $\in$ [0, numUsers]\\

$\vec{bw}[users] = 0  \qquad \forall$ users $\in$ [0,{numUsers}]\\

\While{user interactions exist}{

   \While{no user buffer gets bufferSize new elements}{
        \textcolor{blue}{// loop iterations can execute in parallel}\\
        j $\gets$  {getUserIdFromInteraction()} \\
        $\vec{w} = {Mw[{j}]}^{-1} \times \vec{bw[{j}]}$\\
        choice $\gets$ UCB($\vec{w}, occ[{j}], getContext(j), {Mw[{j}]}^{-1}$)  \\
        ${reward} \gets getReward(choice)$\\
        $update \gets  {context[choice]} \cdot {context[choice]}^\top$\\
        ${Mw}_{local}[{j}] +=  update$\\
        $\vec{bw}_{local}[{j}] +=  {reward} \times context[choice]$ \\
        {Mw[{j}]} += {MBuff}[{j}].pop()\\
        $\vec{bw}[{j}] += {bBuff}[{j}].pop()$\\
        {MBuff}[{j}].push(update)\\
        {bBuff}[{j}].push(reward $\times$ context[choice]\\
        {occ[{j}]} $\gets$ {occ[{j}]} + 1\\        
    }
    shuffledUserOrder $\gets$ {getRandomizedOrder(numUsers)} \\
     \For{each user i}{
        {peer} $\gets$  {getRandomPeer(shuffledUserOrder[i])}\\
        $\vec{w} \gets  {{Mw}_{local}[i]}^{-1} \times \vec{bw}_{local}[i]$\\
        $\vec{v} \gets  {Mw}_{local}[{peer}]^{-1} \times \vec{bw}_{local}[{peer}]$ \\
        \If{|$\vec{w} - \vec{v}$| $\geq \gamma \times {getThreshold}(peer,i)$}{
            remove E[i,peer] \\
            Reset all entries of {MBuff}[i], {MBuff}[{peer}], {bBuff}[i], {bBuff}[{peer}]\\
            Reset {Mw}[i], $\vec{bw}$[i], Mw[peer], $\vec{bw}$[{peer}]\\

        }
        \uElseIf{peer and i share same neighbors}{
            {MBuff}[i] $\gets$ ({MBuff}[i] + {MBuff}[{peer}] ) /2\\
            {bBuff}[i] $\gets$ ({bBuff}[i] + {bBuff}[{peer}] ) /2\\
            Mw[i] $\gets$ (Mw[i] + Mw[{peer}] ) /2\\
            bw[i] $\gets$ (bw[i] + bw[{peer}] ) /2\\
        }

    }
}
\caption{Distributed Clustering Confidence Ball (DCCB) : performs clustering in a distributed manner.}
\label{algo:dccb}
\end{algorithm}

\subsection{Distributed Clustering Confidence Ball (DCCB)}
Listing~\ref{algo:dccb} shows the DCCB \cite{korda2016distributed} algorithm. DCCB is a decentralised, distributed variant of the CLUB algorithm.


DCCB allows catering to different user interactions in parallel. Initially, similar to CLUB, all users are connected to each other.
For each user, DCCB maintains a fixed-sized buffer of interactions-ordered correlation matrices, and $b$-vectors. All elements in the buffers are initialized to zero. Furthermore, each user has an active and a current correlation matrix and a bias-vector associated with it. For each user $i$,  current objects are used to compute the user-vector, i. e.,  $\vec{w}[i]=\textit{Mw}[i]\times\vec{\textit{bw}}[i]$. Each interaction updates the active and the current correlation matrix and the $b$-vector by dequeuing the corresponding entries at the head of the queue. Correspondingly, a new entry of both the correlation matrix and the $b$-vector are inserted into the respective buffers. 

The above parallel user interaction process continues until one of the users has observed an interaction count of \textit{bufferSize}. This stage marks the beginning of network updates. Each user obtains the buffers from a randomly selected user it is connected with. If the user vectors of the two differ by more than a threshold, then the edge is removed, potentially placing the two users into different clusters. Otherwise, if the neighbor sets of the two users match, then the buffers of the user are updated with average values of the peer's buffers, and its own. 
After this cluster formation stage is over, user interactions are addressed in parallel again. Note that unlike CLUB, DCCB does not actively recommend on the basis of a cluster's information.

Unlike CLUB, which updates network per user interaction, DCCB updates the network lazily. It therefore needs to archive the history via buffers. DCCB can be considered as having two repeating stages. The first stage is a massively parallel stage that handles user interactions, while the second stage performs cluster updates. The cluster formation stage requires peer-to-peer communication, although each user only communicates with a single peer.

\begin{algorithm}
\SetAlgoLined
\fontsize{8}{10.9}\selectfont
\KwIn{$\beta$ cluster penalizing hyperparameter}
\KwIn{$\sigma$ initial round splitting hyperparameter}
\KwIn{$G(V,E):$ Graph where vertices represent users. Edges represent correlation between users. Initially it is a complete graph. }
\KwIn{$occ$: Array of counters for number of interactions with each user, initialized to 0}
\KwIn{context: table containing features for all items}
\KwIn{uRounds, cRounds: arrays representing interaction round limits per user for user-based and cluster-based recommendations}

\textbf{Initial state:}
{Mu}[user]= I $\qquad \forall$ users $\in$ [0, {numUsers}] \\
$\vec{bu}$[user] = 0  $\qquad \forall$ users $\in$ [0, {numUsers}]\\
uRounds[users] = cRounds[users] = $\sigma \quad \forall$ users $\in$ [0,numUsers]\\

\While{user interactions exist}{
  \While{ user interactions $\leq$ uRounds[user] $\forall$ users }{
    \textcolor{blue}{// explicitly parallel}\\
    {userID} $\gets$ ${getUserInteraction()}$ \\
    $\vec{u} \gets {Mu[{userID}]}^{-1} \cdot \vec{bu}[{userID}]$\\
    $choice \gets {UCB}(\vec{u}, occ, context(t), {Mu[{userID}]}^{-1}$)  \\
    $reward \gets {getReward(choice)}$\\
    {Mu[{userID}]} += {context[choice]} $\cdot {context[choice]}^\top$\\
    $\vec{bu}$[{userID}] += reward $\times$ context[choice] \\
    occ[{userID}] += 1\\
  }
  {UPDATENETWORK()}\\
  
  {clusters} $\gets$ {recomputeClusters(G)}\\
   \For{each cluster c  in clusters} {
    \For{each user n in c}{  
      Mc[c] += Mu[n]\\
      $\vec{bc}$[c] += $\vec{bu}$[n]\\
    }               
  }
  \While{user interactions $\leq$ cRounds[user] $ \forall$ users} {
      \textcolor{blue}{// interactions across clusters can execute in parallel}\\
      {userID} $\gets$ ${getUserInteraction()}$ \\
      {clusterID} $\gets$ {clusters[userID]}\\
     
      \If{$occ[{userID}] \geq \beta \times  \frac{{numClusterInteractionsSeen}}{{numUsersInCluster}}$ \\}{
        $\vec{\theta}$ $\gets$ $Mu[{userID}]^{-1}$ $\cdot$ $\vec{bu}$[{userID}];
      }
      {
        $\vec{\theta}$ $\gets$ ${Mc[{clusterID}]}^{-1}$ $\cdot$ $\vec{bc}$[{clusterID}];
      }
      {choice} $\gets$ {UCB}($\vec{\theta}$, occ, context(userID), ${Mc[{clusterID}]}^{-1}$) \\
      reward $\gets$ getReward(choice)\\
      {Mu[{userID}]}  += {context[choice]} $\cdot$ {context[choice]}$^\top$\\
      $\vec{bu}$[{userID}]  += reward $\times$ context[choice] \\
      occ[{userID}] +=  1\\
      numClusterInteractionsSeen[clusterID] +=1\\
      
  }
  \For{each cluster c  in clusters} {
      meanOccInCluster $\gets$ $\frac{numClusterInteractionsSeen[c]}{numUsersInCluster}$\\
      \For{each user n in c}{
        $\delta$ $\gets$ $\frac{occ[n] - meanOccInCluster}{\beta}$\\
        uRounds[n]  +=  $\delta$ \\
        cRounds[n] -= $\delta$ \\
      }               
  }

}

\caption{{\name} : Performs clustering lazily, and uses both user-driven and cluster-driven recommendations.}
\label{algo:dclub}
\end{algorithm} 

\section{Design of \name : a highly distributed online recommendation system}
\label{sec:design}

\begin{figure*}[!htb]
\minipage{0.5\textwidth}
  \includegraphics[width=0.9\linewidth, frame]{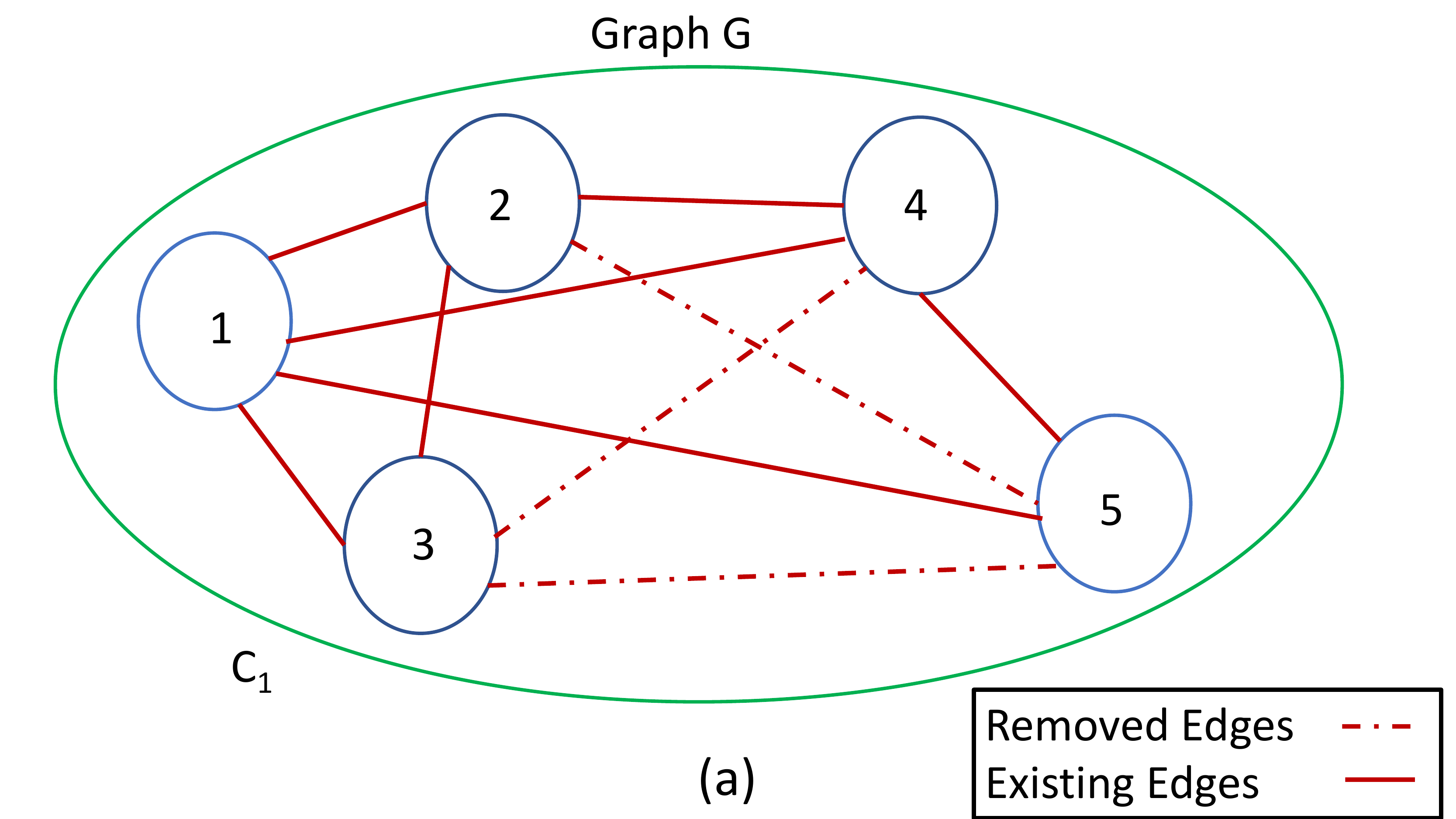}
\endminipage\hfill
\minipage{0.5\textwidth}
  \includegraphics[width=0.9\linewidth, frame]{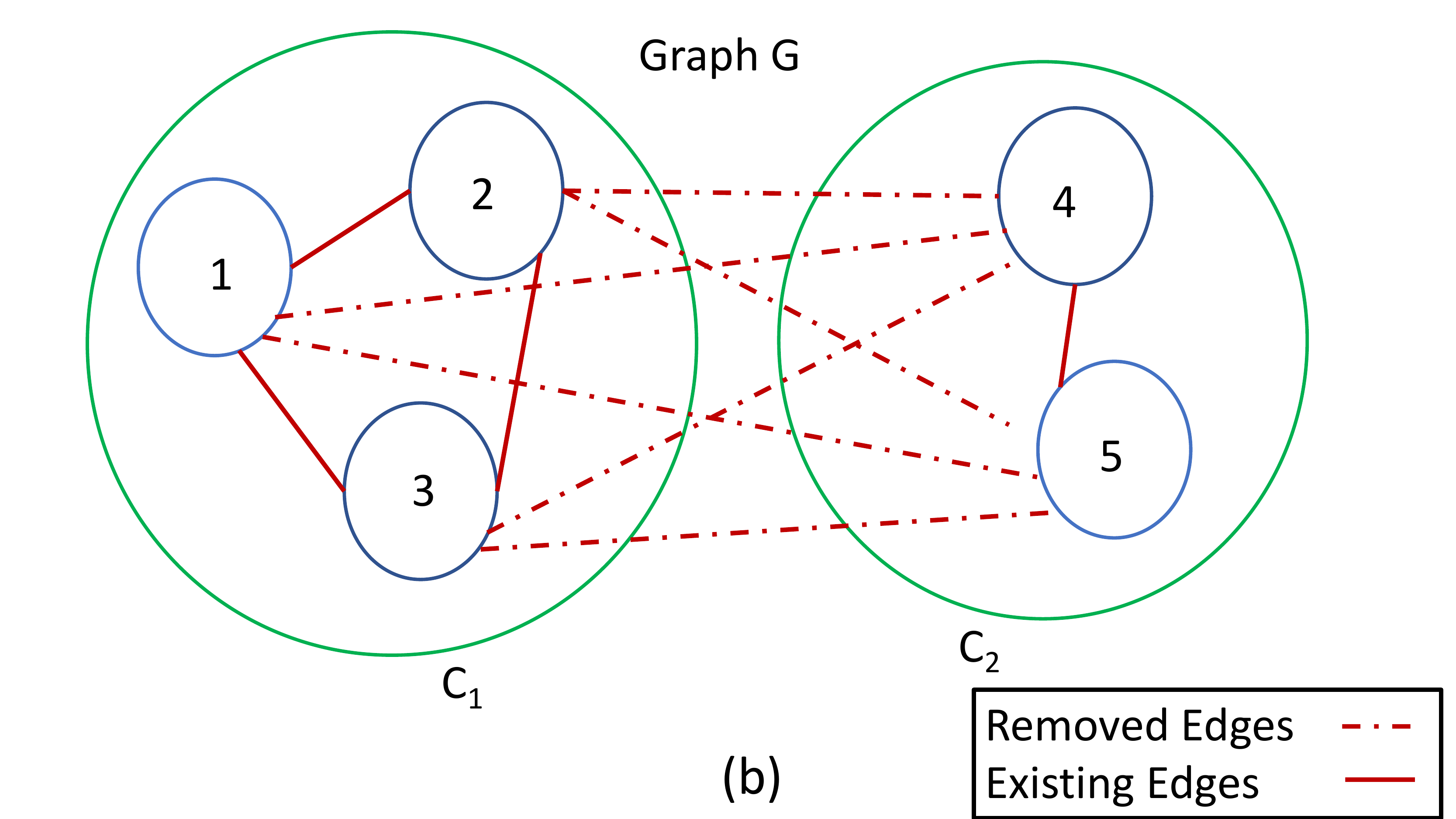}
\endminipage
\caption{Comparison of clustering in DCCB and {\name}. Graph $G$ is made of $|V|=5$ vertices $\{1,2,3,4,5\}$ representing users.. Clustering in DCCB proceeds lazily - hence fewer edges are removed at a given time, as shown in Figure~\ref{fig:clusteringdccbdclub}(a), where edges between users $(1,4)$,$(2,4)$, and $(1,5)$ have not yet been broken. However, in {\name} Figure~\ref{fig:clusteringdccbdclub}(b), clustering is performed on the entire graph $G$, forming the clusters more quickly. This allows parallel interaction processing  on clusters $C_{1}$ and $C_{2}$.}
\label{fig:clusteringdccbdclub}
\end{figure*}

This section describes limitation of DCCB. It presents the {\name} algorithm, and explains how the algorithm overcomes the limitation faced by DCCB. The section also describes the Spark-based {\name} system design.

\subsection{Limitations of DCCB}
DCCB, the state-of-the-art distributed online recommendation algorithm achieves accuracy similar to that of CLUB, but faces the following key challenges prohibiting it from achieving high performance.

\begin{enumerate}

\item Cluster formation requires a global understanding of the graph. DCCB's lazy update-based approach where a node communicates with a single peer to share updates fails to gather the global view quickly. This slows down the cluster discovery, lowering accuracy.

\item DCCB approach archives user-interaction information history in buffers, and communicates the entire history with peers. This results in communication becoming the bottleneck.


\end{enumerate}

The {\name} algorithm overcomes these issues. 

\subsection{{\name} Algorithm}
We continue with the notation introduced in Section~\ref{sec:algos}. 
Listing~\ref{algo:dclub} shows the {\name} algorithm. {\name} employs three main strategies
a) {Efficient parallelism exploitation} : When recommendations do not use clustering information, and are based on the user's past interactions alone, user-level interaction parallelism can be exploited explicitly. However, good recommendations require clustering information too. As clusters are discovered, updates to a cluster's information must occur sequentially in order to maintain consistency. A key principle of {\name} algorithm is to interleave stages of user-based and cluster-based interactions. In the latter, although interactions of users in a given cluster cannot be parallelized, interactions \textit{across} clusters can occur in parallel. 
b) {Limiting communication bottleneck} :
{\name} uses no buffers. Instead, it only communicates each user's correlation matrix and user-vector, each time clustering is updated.
c) {Adaptive user-based and cluster-based round limits} : {\name} dynamically switches between using cluster-specific information, when information for a user is sparse, and highly personalized user-specific information, when information for the user is abundant. This allows {\name} to achieve high accuracy.

To realize the above strategies we design {\name} algorithm to have four major repeating stages described as follows:

\subsubsection{Stage 1 : Recommending via user information alone}

Initially, user correlation matrices and bias vectors contain very little information. However, in subsequent iterations of the algorithm, the reward-based feedback mechanism updates the user correlation matrix and the user-vector. 
User-vector updates facilitate two outcomes:
i)Improved recommendations as more user information is gathered, and
ii) Clusters discovery.
The recommendations are generated using the linear contextual bandit setting, similar to CLUB. Since users share no data elements, interactions in this stage are embarrassingly parallel. Note that interactions of a given user are processed in order.  
Cluster updates are deferred to the third stage of the algorithm. 



\subsubsection{Stage 2: Network updates and cluster formation}
This stage is logically equivalent to the network update stage in CLUB. It first updates the network by removing unnecessary edges. Next, it generates clusters and computes their correlation matrices and cluster-vectors. This stage can exploit tree-reduction style parallelism, as will be described in Sec.~\ref{sec:impl}.

\subsubsection{Stage 3: Cluster-based recommendations}
Figure~\ref{fig:clusteringdccbdclub} presents a comparison of how clustering occurs in {\name} and DCCB. The faster cluster formation in {\name} creates an opportunity for generating cluster-specific recommendations.

This stage is similar to stage 1, except that instead of using the user's information for recommendations, cluster's information is used.  This helps normalize the recommendations across all users in the cluster. A distinguishing aspect of this stage is that for users with significantly higher interaction counts than other users in the cluster, the recommendation is still provided on the basis of the user's information, making the recommendation more personalized. This is crucial because in spite of the clustering, users with an exceptionally long history are more likely to choose an item based on their own past activity. To identify such users, {\name} computes the number of interactions processed for the user, and determines if it is $\beta$ times more than the average interactions for users in the cluster. If so, then the individual user's information is utilized for recommendations, else cluster information is used. Interactions where the catered users belong to different clusters can be processed in parallel in this stage.

\subsubsection{Stage 4: Balancing stage 1 and 2 sizes}
For users who have observed more interactions than others in a given cluster, personalized recommendations will be more accurate. On the contrary, for users with few interactions, recommending based off cluster's information offers better rewards. 
The number of user interactions to be processed according to cluster information and user information are dynamically chosen in {\name}. 
Initially, stage 1 and 2 operate on \textit{uRounds} and \textit{cRounds} arrays respectively, which are initialized to the same values for all users.
This stage factors in information regarding interaction volume for each user, and updates user and cluster round limits accordingly, by a factor $\delta$, as shown in Algorithm~\ref{algo:dclub}.
This mechanism automatically steers newly coming users towards using cluster-based recommendations.
This approach makes {\name} a hybrid of CLUB and a fully personalized linear bandit approaches, with the decision to switch taken in an adaptive fashion.

\subsection{Discussion}

\begin{figure*}[!htb]
\minipage{0.5\textwidth}
  \includegraphics[width=\linewidth, frame]{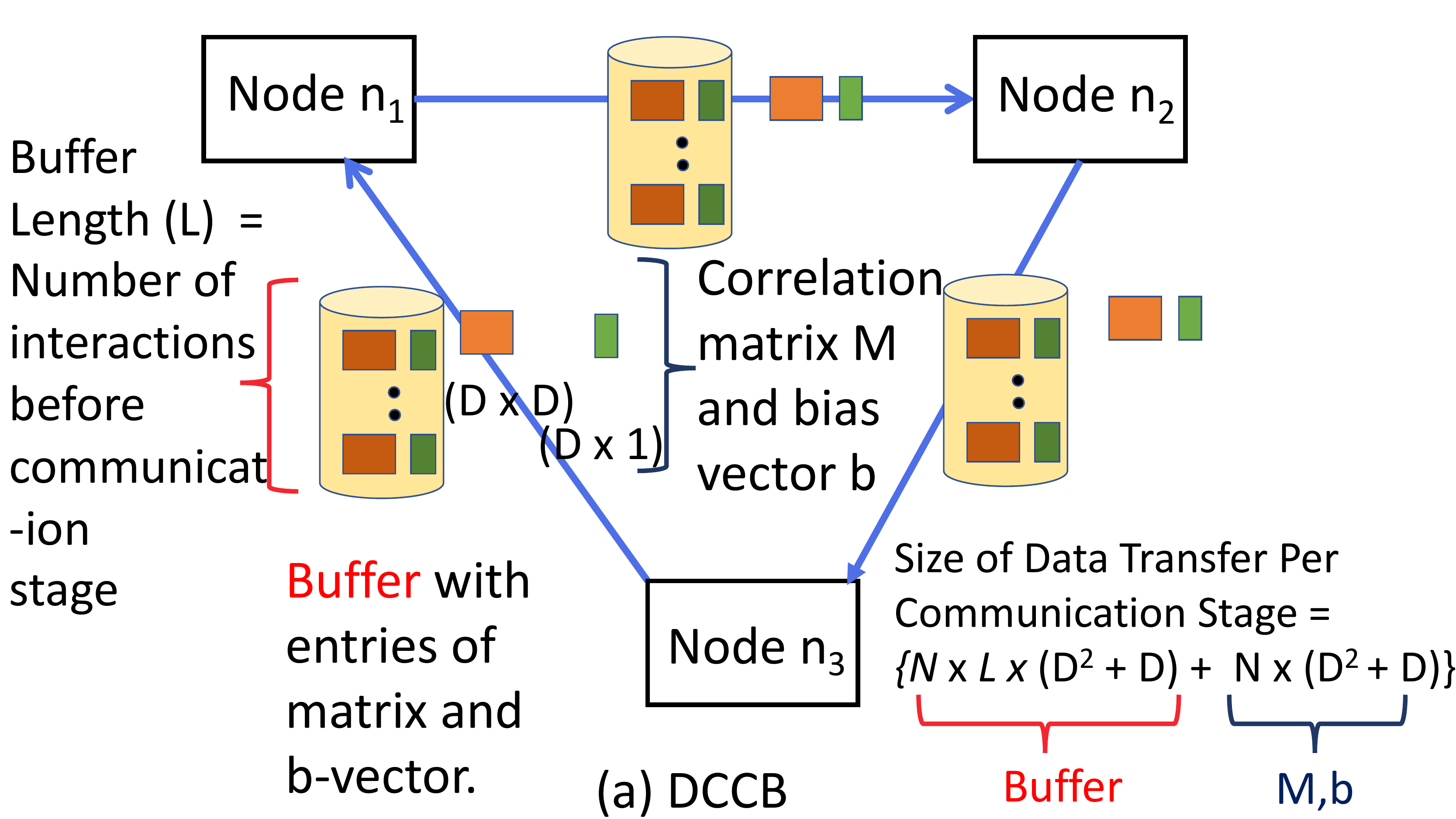}
\endminipage\hfill
\minipage{0.5\textwidth}
  \includegraphics[width=\linewidth, frame]{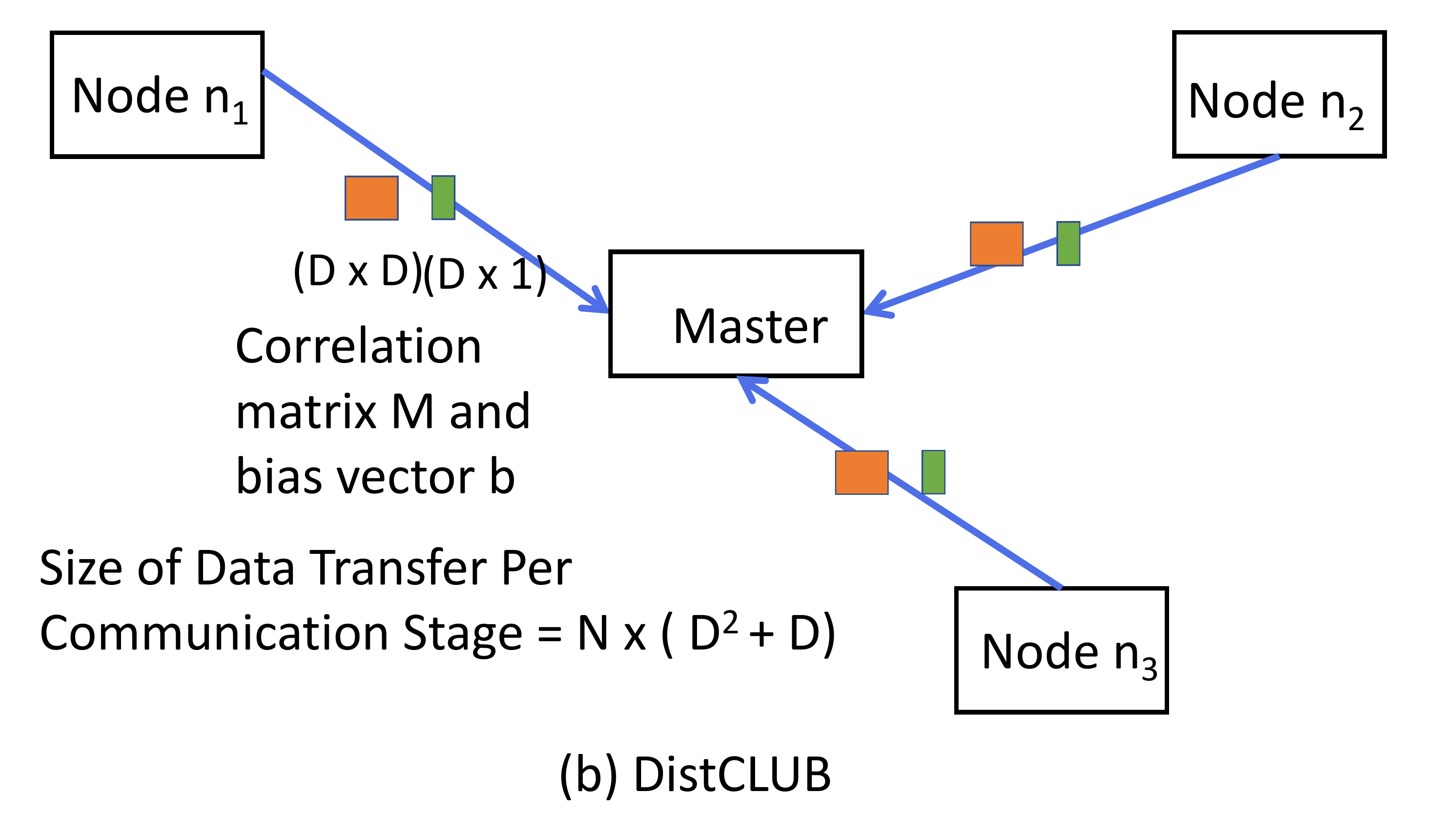}
\endminipage
\caption{Comparing communication requirements: This figure shows a scenario where each node caters to a single user. $N$ is the total number of nodes,  $D$ is the feature count. Figure 2(a)- In DCCB each node maintains a buffer of length $L$ which holds an archive of user matrices for each interaction, along with the active matrices. The buffers result in large communications. On the other hand, data exchange in {\name} is much lighter (Figure 2(b)). Table~\ref{tab:info} shows cumulative data transfer sizes for DCCB and {\name} required in practice, where each node caters to several users.} 
\label{fig:dccbvsdclub}
\end{figure*}

While both DCCB and {\name} are distributed online recommendation algorithms where nodes share information, there are important differences in their design. Figure~\ref{fig:dccbvsdclub} compares the data communication sizes  of DCCB and {\name}. In DCCB, a user shares the archived buffer with a random peer. This partial information sharing results in biasing, and only gets mitigated when several communication rounds have occurred. Design alternatives to reduce DCCB data transfers are 
i) Reducing buffer lengths - if buffer lengths were to be reduced, frequency of communication will increase, amplifying the associated communication overheads, while keeping the overall communication sizes the same.
ii) Reducing the number of users that share information in communication round - This approach will maintain the frequency of communication and buffer lengths, but will lower the overall communication volume. Unfortunately, this results in accuracy reduction. Both these alternatives are thus rendered ineffective.  The communication bottleneck exacerbates in environments with large user counts because each machine must handle several users, constraining data transfer limits. 
On the other hand, 
{\name} aggregates and processes information from all nodes during a communication stage removing bias in information sharing.

\subsection{Implementation and Optimizations}
\label{sec:impl}

\begin{figure}[!htb]
\includegraphics[width=80mm,height=55mm]{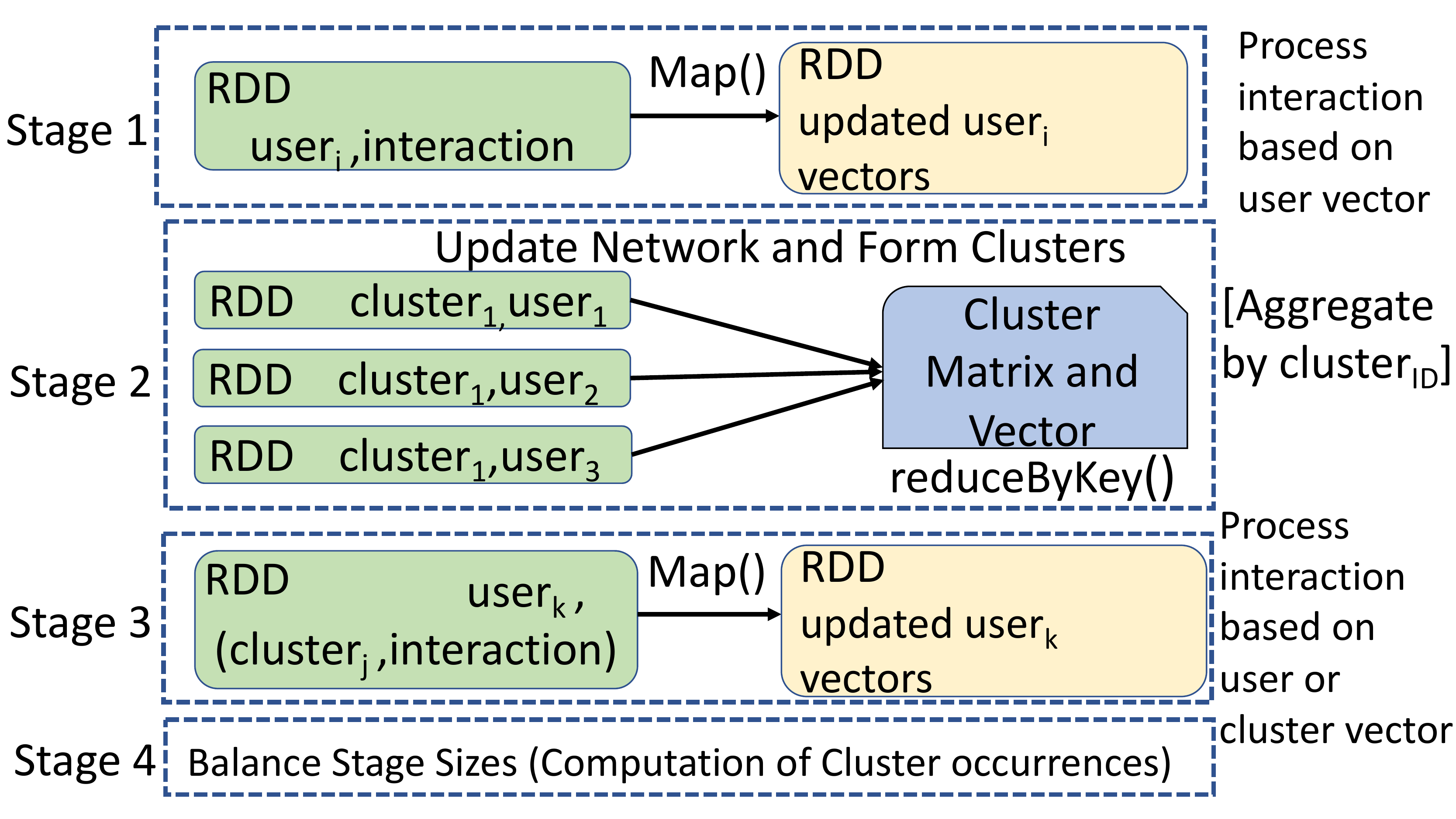}
\caption{{\name} System Implementation in Spark. Figure depicts RDDs in {\name}'s Spark Jobs.}
\label{fig:flow}
\end{figure}

\textbf{Choice of Framework}
We chose to use the Spark Framework~\cite{zaharia2016apache} to implement {\name}, owing to its support for large-scale distributed processing, fault-tolerance, in-memory data persistence, and automatic data locality exploitation. Furthermore, Spark automatically exploits both inter-node and intra-node parallelism. Other traditional frameworks such as MPI do not provide automatic in-memory data persistence and out-of-the-box data locality exploitation like in Spark. Spark is highly popular, easy to use, and well-integrated with other big-data processing frameworks. It also supports parallel graph processing libraries.  Dask~\cite{dask} could be a handy framework to parallelize such applications and is lighter weight than Spark, however it is not yet adapted in the Big Data enterprise world.

Spark uses a master-worker paradigm where the Spark Executors are the workers and they execute Map tasks in parallel. Spark jobs have an Application Master that controls launches for all tasks. 

Figure~\ref{fig:flow} depicts the mapping of {\name} algorithm on Spark. We represent the user correlation matrix M, the vector b, the user-graph G, the cluster correlation matrix M and the   and the user vectors as RDDs (Resilient distributed datasets)~\cite{zaharia2010spark}. RDD is a Spark primitive, representing a fault-tolerant collection of elements that can be operated on in parallel. Additionally, we make Spark persist these RDDs in the node's memory after each update to them. This way, a subsequent interaction for the same user is fast. This is primarily because of Spark's scheduling mechanism that tries to allocate next such task to the same node.  Spark replicates RDDs across multiple nodes in the cluster for fault-tolerance thus making {\name} robust.
Spark Map transformation applies a function to each element in the RDD in parallel and returns a new RDD. We leverage the Map transformation to process the user interactions with the key-value pair being $<$userID, user-interaction to process$>$ for stage 1 and with a slight modification in stage 3 of {\name}. 

For both these stages, if there are multiple interactions with the same user, their execution is serialized.
Spark automatically schedules and executes these Map tasks such that the data which they process is available in the local disk, and preferably in the memory of the node. This typically results in tasks for a given user being executed on the same set of machines. The updated user correlation matrix and b-vector are passed back to the Spark Master. 

For Stage 2 of {\name}, to process network updates efficiently, we perform the edge updates for each user in parallel Map tasks. This merely involves communicating the user-vector information, which is small in size. Clusters are generated based on the updated edges; by performing computation of \textit{connected components} in a distributed and parallel manner in Spark~\cite{xin2013graphx}. Updated user vectors are computed using the updated correlation matrix and b-vector. Cluster vectors are calculated for the newly formed clusters. This is done by summing up the constituent user vectors in each cluster. To prevent this step from becoming a bottleneck, aggregation of user-vectors into the cluster vectors is performed in parallel, for each cluster. To this end, we make use of Spark's tree aggregation pattern, and utilize treeReduce to collect the correlation matrix and b-vector of users in a cluster.

Stage 4 of the algorithm is trivially parallelizable using Map tasks in Spark. In this manner, all four stages of {\name} are \textit{optimized} and mapped into \textit{efficient} distributed and parallel primitives in Spark.

\section{Analysis of {\name}}
\label{sec:analysis}

The regret of a bandit learner relative to a learning policy is defined to be the difference between the total expected reward using that policy for $T$ recommendation rounds and the total expected reward collected by the learner over $T$ rounds. Therefore, regret analysis is used to compare the learning rate of {\name} with alternative approaches, such as DCCB and CLUB. This section shows that regret bounds for  {\name} are similar to those of DCCB and CLUB. To derive the regret bounds, we first establish our learning model.

\subsection{Learning model for {\name}}
 Let $V =
\{1, . . . , n\}$ represent the set of $n$ users. User behaviour similarity induces a partitioning of V into $c$ sets $V_{1}$, $V_{2}$,..., $V_{c}$ with $c << n$.  Users belonging to the same set or cluster share similar behavior while users from different clusters have different behaviors. Note that, some sets in these partitioning can be singleton sets, such that the user does not share similar behaviour with other users.
No partitioning information of $V$ or user behavior are known to the learner in the beginning. In each user interaction $t$ the learner receives a user $i_{t}$, together with a set of context vectors $C_{i_{t}} \subset R^d$ from which the learner chooses an action $ A(t)$. $d$ represents the number of features for each action in $C_{i_{t}}$. Subsequently, the learner observes reward $W(t)$, based on the chosen action.  We assume $i_{t}$ is selected uniformly at random from $V$.

\subsection{Regret Analysis}
\label{sec:regret}

Denote $T$ as the total number of iterations and $T_{i}$ as the total number of iterations when user $i$ is selected up to time $T$. So we have $T = \sum_{i \in V} T_i$. Standard contextual bandit analysis shows that, for a particular user, as $T_{i}$ grows large, the cumulative regret 
can be bounded with high probability~\cite{abbasi2011improved,gentile2014online} as 
\begin{equation}
R_{\text{IND}}(T,i)
= O(\sqrt{d T_i})
\end{equation}
Where $r_{t,i}$ is the regret from interaction processed at time $t$ with user $i$, and $R_{\text{IND}}(T,i)$ is the aggregate regret over $T$ interactions for user $i$ by running an independent linear contextual bandit model. The total aggregate regret over all the users, denoted as $R_{\text{IND}}(T)$ can be written as, 
\begin{equation}
R_{\text{IND}}(T) = \sum_{i \in V} R(T,i) = O(\sum_{i \in V} \sqrt{d T_i})  
\end{equation}

If users are clustered, which means that observations are shared across users, the regret bound of a cluster of user $V_j$ and over all regret bound can be written as,

\begin{equation}
R_{\text{CLU}}(T,V_j) = O(\sqrt{d \sum_{i \in V_j} T_i})
\end{equation}

\begin{equation}
R_{\text{CLU}}(T) = \sum_{j=1}^c R_{\text{CLU}}(T,V_j) = O( \sum_{j=1}^c \sqrt{d \sum_{i \in V_j} T_i})
\end{equation}

Since users $i_{t}$ are chosen randomly, we can consider $T_i = \frac{T}{n}$, the above regret can be simplified as,
\begin{equation}  \label{eqn:indpendent}
R_{\text{IND}}(T) = O( \sqrt{dT} \sum_{i \in V} \sqrt{1/n})  
\end{equation}

\begin{equation} \label{eqn:cluster}
R_{\text{CLU}}(T) = O( \sqrt{dT} \sum_{j=1}^c \sqrt{|V_j|/n})
\end{equation}

If we consider regret in the clustering case, i.e. $R_{\text{CLU}}(T)$ as the reference regret, regret is only affected when user interactions are processed in {\name}. This happens in Stage 1 and Stage 3 of the algorithm. In the other stages, regret remains unchanged. The regret bound depends on the size of clusters $V_{j}$.  In Stage 1, the user correlation matrix and the user-vector are considered only for choosing the action $A(t)$, and no information is shared across users. In this case, regret is the same as running independent linear contextual bandit models. Hence the regret bound in this stage, denoted as $R_{S1}$, can be written as,
\begin{equation} \label{eqn:degen}
R_{\text{S1}}(T) = R_{\text{IND}}(T) = O(\sqrt{ndT})
\end{equation}

After clustering of users in Stage 2, assume that $m$ clusters are formed such that they contain $n_{1}$ users overall, and let there be $n_{2}$ users ($n_{2}<<n, m<<n$) whose rewards are generated without considering their cluster information. Therefore, $n_{1}+n_{2}=n$ also $m+n_{2}=c$. In the worst case, i.e. make the summation term significant, let $m$ clusters have the same size $n_{1}/m$, Hence substituting this information back in equation(~\ref{eqn:indpendent}) and  equation(~\ref{eqn:cluster}), we have,
\begin{align}
R_{\text{S2}}(T) & = O(\sqrt{dT} ({m}\times\sqrt{(n_{1}/m)/n}+n_{2}\times\sqrt{1/n})) \\ \nonumber
& = O(\sqrt{dT} (\sqrt{n_1 m/n}+\sqrt{n_2^2/n})) 
\end{align}
Since $n_{2}<<n, n_{1}/n =~1$ and $n_2^2/n$ can be dropped because of the O notation.
Hence
\begin{equation}
R_{\text{S2}}(T) = O(\sqrt{dTm})
\end{equation}

In the degenerate case, the regret will fall back to equation (\ref{eqn:degen}).
The regret thus has the same bounds as DCCB and CLUB, and as the number of user interactions processed by {\name} tends to $\infty$, the final regret is asymptotically equivalent to CLUB and DCCB~\cite{gentile2014online,korda2016distributed}. 
\section{Evaluation}
\label{sec:eval}
In this section we present an experimental comparison of {\name} with DCCB and CLUB. The experiments compare execution times and prediction performance of {\name}, CLUB, DCCB for multiple datasets. The section also compares scalability of {\name} and DCCB.

\subsection{Experimental Setup}


We used Amazon EC2 to run our experiments. Each compute node is a c5.4x~\cite{c5i} instance with 32 GB of RAM. Each node contains eight cores. The c5 instance has an Intel Xeon Platinum 8000 series processor with CPU clock speed of up to 3.6 GHz. The machines in this cluster have a network bandwidth of up to 10 Gbps. 

\begin{table}[!htbp]
  \begin{center}
    
    \begin{tabular}{|c|c|c|c|} 
    \hline
     \textbf{Dataset} & \textbf{Number of } & \textbf{Number}&\textbf{Number of}\\
     & \textbf{User Interactions} & \textbf{Of Users} & \textbf{Item Features} \\ 
      \hline
      MovieLens & 80,000 &  943 & 19 \\
      LastFM    &  10,000     & 1,888 & 25\\
      Delicious&  10,000       &1,816 &25\\
      Yahoo &50,000 & 5,045 & 1  \\
      Synthetic & 4,000,000 & 20,000 & 25 \\
\hline
    \end{tabular}
\caption{Datasets used in the evaluation}
\label{table:dataset}
  \end{center}
\end{table}

We used Spark version 2.4.3 and EMR version 5.26.0 in the experiments. The EMR cluster was run in cluster mode, wherein the Application Master and Spark driver together with the Spark Executors occupy all cores in the system.

Table~\ref{table:dataset} shows the datasets we used for the experiments. The values used for all configurable parameters are shown in Table~\ref{tab:parameters}. We followed the dataset preparation described in Li et al.~\cite{gentile2014online}. These datasets are used to evaluate {\name} in dynamic environments where the set of active users and user preferences change rapidly, and have been used in past for evaluating CLUB and DCCB algorithms~\cite{gentile2014online,korda2016distributed}. The evaluation is performed over four real-world benchmark datasets and one synthetic dataset, each having different user, interaction, and item feature counts. The MovieLens dataset~\cite{lam2006movieLens} contains personal ratings and tags for multiple users.
Delicious is a social bookmarking webservice system, while LastFM is a social music streaming system. These datasets were obtained from ~\cite{datasets}. The Yahoo dataset is obtained from the ICML 2012 Exploration and Exploitation Challenge for news article recommendation and was preprocessed following the directions in ~\cite{gentile2014online}. The synthetic dataset was created with twenty thousand users and four million user interactions. We created the synthetic dataset to stress {\name} with much larger number of users, clusters, and user-interactions than the above datasets. The synthetic dataset models today's recommendation systems that have high volume of users and are expected to process interactions at a high rate. Table~\ref{tab:parameters} shows the various configurations/hyperparameter values used in the evaluation.

\begin{table}[!htbp]
  \begin{center}
    \begin{tabular}{|c|c|} 
    \hline
    \textbf{Parameter} & \textbf{Value}\\
    \hline
    $\alpha$ : Exploration Parameter in UCB & 0.03\\
    $\beta$ : Cluster penalizing parameter in {\name} & 2\\
    $\gamma$ : Cluster Exploration Parameter in CLUB & 0.7 \\
    $\delta$ : networkUpdate Delay for CLUB & 2000 \\
    \textit{bufferSize} : Buffer length for DCCB & 5000 \\
    $\sigma$ : Initial split sizes in {\name} & 2500 \\
 \hline
    \end{tabular}
 \caption{Hyper-parameter values used in the experiments}   
 \label{tab:parameters}
 \end{center}
 \vspace{-2mm}
 \end{table}

\textbf{DCCB Implementation:}
Since no open-source system implementation for DCCB is available, we implemented the DCCB algorithm (Listing~\ref{algo:dccb}) in Spark, by closely following the description in their paper~\cite{korda2016distributed}. In DCCB, interactions of different users are processed in parallel. We used Spark Map transformation to implement this stage, where the key-value pair is $<$userID, user-interaction to process$>$. After this stage, information sharing happens in a peer-to-peer manner. Note that big data processing systems such as Hadoop or Spark do not directly allow processes to share information outside of their reduction primitives. To facilitate this information sharing to occur in a parallel fashion, we created a Spark Map transformation where the key-value pair is $<$receiving-userID, (sender-ID, senderM, senderb, senderMBuffer, senderbBuffer)$>$. Inside this transformation, the receiving user's buffers are active objects and are updated according to the algorithm. 

\subsection{Execution Time Performance Comparison}
\begin{figure}[!htb]
\centering
\includegraphics[width=\linewidth,frame]{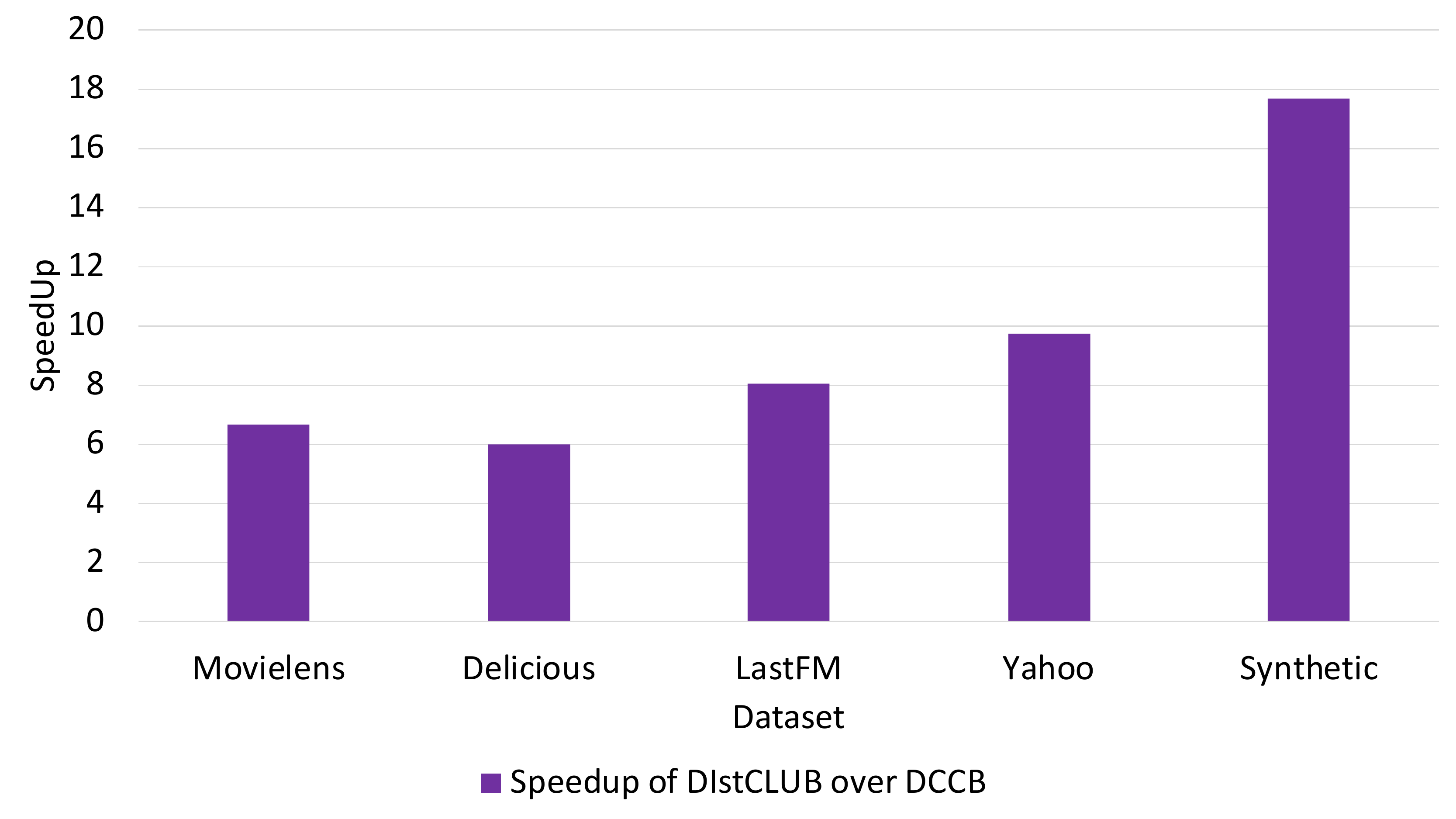}
\caption{Speedup of {\name} over DCCB on  different datasets. The geometric mean speedup is {\perf}.}
\label{fig:speedup}
\end{figure}

Table~\ref{tab:exec} shows the execution times in seconds for CLUB, DCCB, and {\name}. CLUB cannot exploit cross interaction parallelism, while DCCB and {\name} exploit interaction-level parallelism executing on 8 nodes, where each node has 8 cores. CLUB can exploit intra-interaction parallelism, such as in the matrix multiplication routines. Figure~\ref{fig:speedup} plots the speedup achieved by {\name} over DCCB. We could not finish running the synthetic dataset in CLUB because it did not finish within three hours. We executed each run thrice, and calculated the average execution times. {\name} outperforms both DCCB and CLUB. The geo-mean speedup achieved by {\name} over DCCB and CLUB is 8.87x and 29.13x respectively. For the synthetic dataset, {\name} runs 17.65 times faster than DCCB because communication overheads become severe with large user and interaction counts.

\begin{table}[htbp]
  \begin{center}
    \begin{tabular}{|c|c|c|c|} 
    \hline
    \multicolumn{4}{|c|}{Execution Time in seconds} \\
    \hline
    \textbf{Dataset} & \textbf{CLUB}& \textbf{DCCB}& \textbf{{\name}}\\
    \hline
    MovieLens & 2395 & 452.5 & 68\\
    LastFM & 1459 & 294 & 49\\
    Delicious & 2074 &  249 & 31\\
    Yahoo &  1785 & 1693 &174 \\
    Synthetic & - & 7730 & 438 \\
 \hline
    \end{tabular}
\caption{Execution times of {\name}, CLUB, and DCCB in seconds. DCCB and DistCLUB use 64 cores (8 nodes). CLUB does not exploit interaction-level parallelism. 
}  \label{tab:exec}
 \end{center}
 \vspace{-2mm}
 \end{table}

\begin{figure}[h]
\centering
\includegraphics[width=\linewidth,frame]{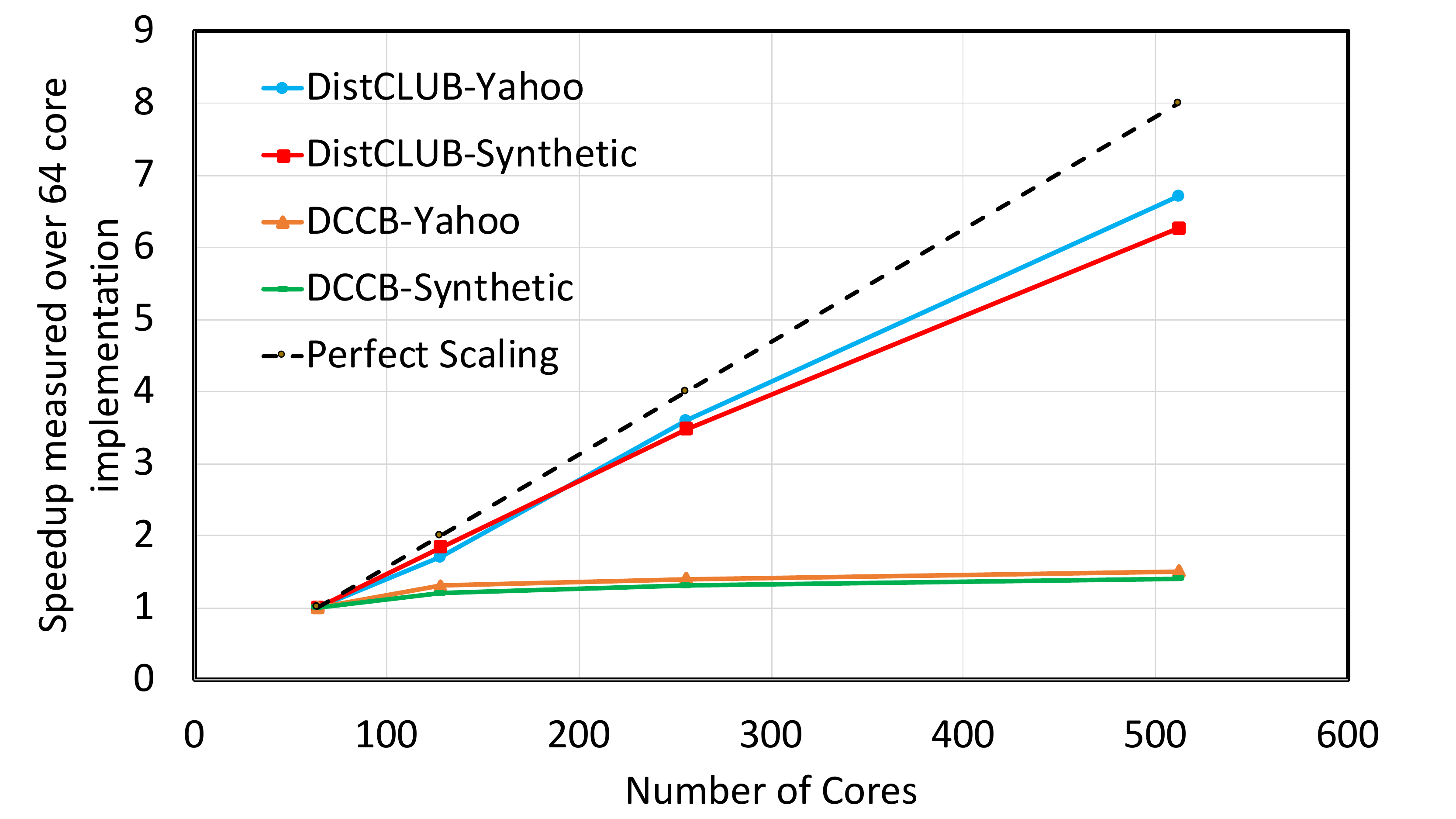}
\caption{Scalability comparison of {\name} and DCCB : DCCB does not scale beyond 128 cores owing to communication overheads.}
\label{fig:scalability}
\end{figure}

\subsection{Scalability}
To evaluate the scalability of {\name} and DCCB we ran them on a real dataset (Yahoo) and a synthetic dataset.  
The synthetic dataset has greater number of interactions and users than any other dataset.

We increased the number of cores from 64 to 512 (going from 8 to 64 nodes) and measured the speedup achieved, compared to the 64 core variant. {\name} has an average scaling efficiency~\cite{scalingefficiency} of 81.2\%. This shows that {\name} can effectively leverage all cores in the system and scale to larger number of nodes.

For large user and interaction counts, each node in the distributed setup ends up catering to several users. In DCCB, this results in the reduction of effective memory available to each user. For large user and interaction counts, buffer sizes shrink significantly, resulting in more frequent communication rounds.

\subsection{Size of information passed around in {\name} and DCCB}

Table~\ref{tab:info} shows the cumulative data shared by DCCB and {\name} during the execution of various datasets over the total interactions. DCCB shares significantly more data than {\name}. The data transferred in DCCB during each communication stage comprises both the buffer and the active objects, and the overall data size depends on the 
number of interactions and buffer length (see Figure~\ref{fig:dccbvsdclub}). 
On the other hand, in {\name}, data transfer is driven by the interaction count alone.

\begin{table}[!htbp]
  \begin{center}
    \begin{tabular}{|c|c|c|c|} 
    \hline
    \textbf{Dataset} & \textbf{DCCB} &  \textbf{\name} \\
    \hline
    MovieLens & 114.6 GB & 29.9 MB\\
    LastFM & 46 GB & 11.1 MB \\
    Delicious & 42.2 GB &10.4MB\\
    Yahoo & 2.1 GB & 423 KB\\
    Synthetic & 204 TB & 43.6 GB \\
 \hline
    \end{tabular}
\caption{Comparison of overall data transfer sizes in DCCB and {\name} : Cumulatively, DCCB shares significantly more information than {\name}}  
 \label{tab:info}
 \end{center}
 \vspace{-3mm}
 \end{table}  

\begin{table}[!htbp]
  \begin{center}
    \begin{tabular}{|c|c|c|c|c|} 
    \hline
    \textbf{Dataset} & \textbf{\name} & \textbf{DCCB} & \textbf{CLUB} & \textbf{\name}  \\
     &\textbf{Reward } & \textbf{Reward }   & \textbf{Reward }& \textbf{to DCCB}\\
    &{\textbf{units }}& \textbf{units } & \textbf{units }&\textbf{reward}\\
    \hline
    MovieLens & 5,620 & 5,321 & 7,014& 1.06 \\
    LastFM & 1,242 & 896 & 1,494 &1.39\\
    Delicious & 484 & 456 & 428& 1.06\\
    Yahoo & 2,279 & 2,138 & 2105 & 1.07 \\
     \hline
    \end{tabular}
\caption{Comparison of normalized cumulative rewards units obtained by {\name} in comparison to DCCB and CLUB. The last column highlights {\name}'s prediction performance compared to DCCB for each of the dataset. {\name}'s prediction performance is \accr higher than DCCB normalized over all datasets} 
   \label{tab:rewardcomparison}
 \end{center}
 \vspace{-3mm}
 \end{table}  

\begin{figure}[!htb]
\centering
\includegraphics[width=80mm,height=45mm,frame]{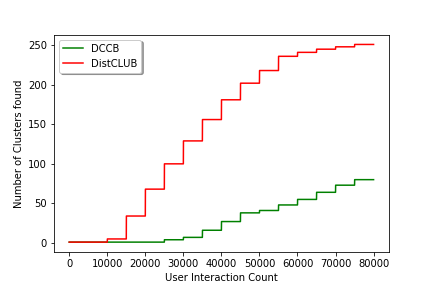}
\caption{Cluster discovery rate in DCCB and {\name}. As interaction count grows, {\name} discovers more clusters than DCCB. Faster cluster discovery results in better recommendations.}
\label{fig:clusd}
\end{figure}

\subsection{Cluster Discovery Speed}
Figure~\ref{fig:clusd} compares the cluster formation rate of DCCB and {\name} on the MovieLens dataset. The figure shows that {\name} finds clusters faster than DCCB. Other datasets show similar patterns in cluster growth rate. Information sharing in DCCB is restricted to a single peer node, while in {\name}, information from all nodes is globally aggregated.
As a result, information sharing is much slower in DCCB, and this slows down the formation of clusters as shown in Fig~\ref{fig:clusd}.

\begin{figure*}[!htbp]
\centering
\minipage{0.5\textwidth}
  \includegraphics[width=\linewidth,frame]{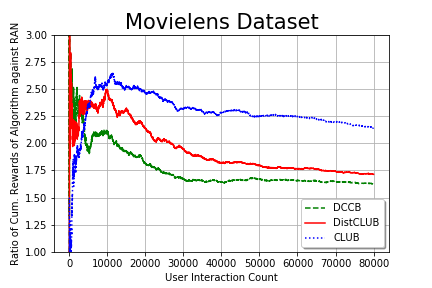}
\endminipage\hfill
\minipage{0.5\textwidth}
  \includegraphics[width=\linewidth, frame]{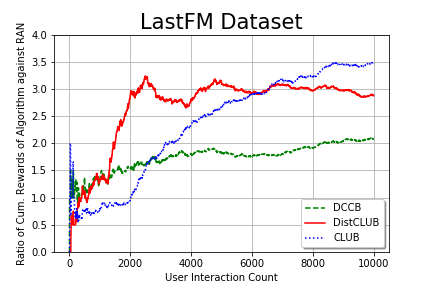}
\endminipage\hfill
\minipage{0.5\textwidth}
 \includegraphics[width=\linewidth, frame]{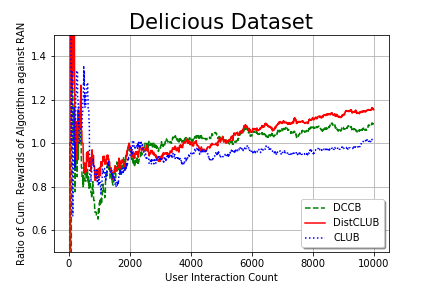}
\endminipage\hfill
\minipage{0.5\textwidth}
 \includegraphics[width=\linewidth, frame]{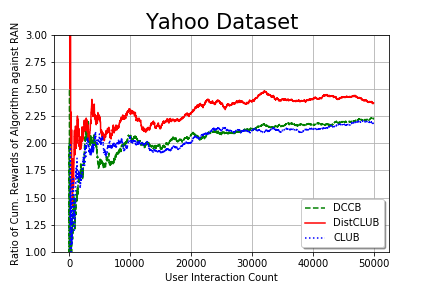}
\endminipage
\caption{ Cumulative reward compared to random action (RAN) for 
{\name} (red), DCCB (green), and CLUB(blue) various datasets: Near the 0 horizon the algorithm behaviour is erratic since  the initial exploration is noisy. Prediction performance (cumulative reward) is higher for {\name} than DCCB for all datasets.}
\label{fig:accuracy}
\end{figure*}

\begin{figure*}[!htbp]
\centering
\minipage{0.5\textwidth}
  \includegraphics[width=\linewidth,frame]{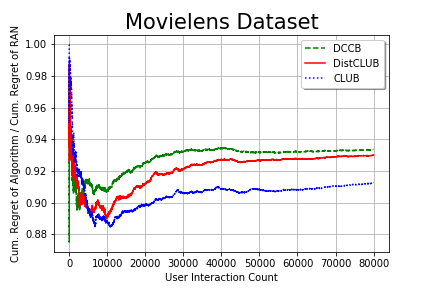}
\endminipage\hfill
\minipage{0.5\textwidth}
  \includegraphics[width=\linewidth, frame]{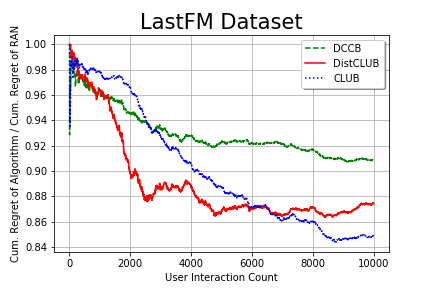}
\endminipage\hfill
\minipage{0.5\textwidth}
 \includegraphics[width=\linewidth, frame]{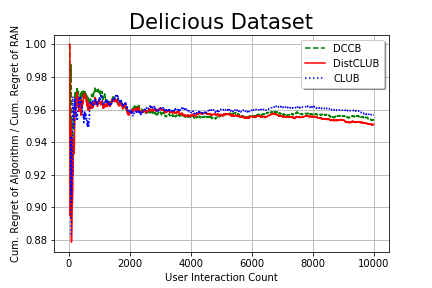}
\endminipage\hfill
\minipage{0.5\textwidth}
 \includegraphics[width=\linewidth, frame]{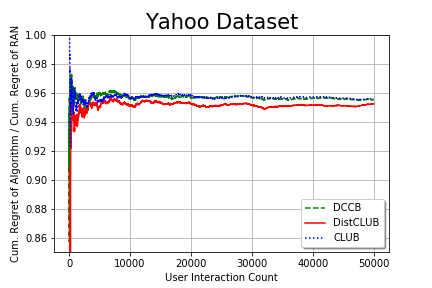}
\endminipage
\caption{ Regret values obtained for various algorithms :
{\name} (red), DCCB (green), and CLUB(blue) for MovieLens, LastFM , Delicious, and Yahoo datasets.   DCCB regret is higher than {\name} for all datasets (lower is better). }
\label{fig:regret}
\end{figure*}

\subsection{Cumulative Reward Comparison}

We measure the ratio of cumulative reward obtained  for \name, CLUB, and DCCB to that obtained by a predictor which chooses an item within a context fully at random
as a recommendation. Figure~\ref{fig:accuracy} shows results of this comparison, where each run was executed thrice. The algorithm behaviour is erratic near the zero horizon. This is because the exploration in the beginning interaction rounds is noisy. The objective is
to achieve largest possible cumulative reward over all user interactions~\cite{lattimore2018bandit}. We divide the cumulative reward obtained by the algorithm with the cumulative reward obtained by a predictor which chooses a random item, and then obtain average prediction performance across datasets.  {\name} achieves {\accr} higher \textit{prediction performance} than DCCB {normalized} over all the datasets. Because reward value is 0 or 1 in all datasets, achieved rewards can be averaged and compared across datasets. Table~\ref{tab:rewardcomparison} shows the performance on each individual dataset.
 
On LastFM and Yahoo datasets, {\name} performs much better than DCCB by efficiently finding more clusters, indicating that social information plays higher role in achieving better predictions on these datasets. CLUB performs the best, owing to the fact that it serially processes individual interactions, finding better clustering.

\subsection{Cumulative Regret Comparison}
Figure~\ref{fig:regret} compares the ratio of the cumulative regret for {\name}, CLUB and DCCB to a random predictor RAN which randomly chooses an item. The cumulative regret is the difference between the algorithm's accumulated reward and the maximum the algorithm could have obtained if all parameters are known, so lower  regret is better. Regret is compared in the same setting as the rewards. For all datasets, overall regret for {\name} is lower than DCCB. The experiment reaffirms the regret analysis presented in Sec.~\ref{sec:regret} which showed that asymptotic regret bounds for {\name} are same as that of DCCB and CLUB.

\section{Related Work}
\label{sec:related}

\textbf{Offline recommendation algorithms}
Recommendation systems' algorithms can be classified into two major methods - Collaborative Filtering (CF) and Content-based methods. CF techniques calculate similarity based on interactions. CF techniques use algorithms like k-nearest neighbors~\cite{peterson2009k} and matrix factorization~\cite{koren2009matrix} to predict the utility of items to users. SGD~\cite{zhuang2013fast} or ALS~\cite{pilaszy2010fast} are used in training such offline algorithms. Content-based recommendation systems recommend items to users based on user's interest profile and item descriptions~\cite{pazzani2007content}. Techniques such as Decision Trees~\cite{cho2002personalized}, Naive Bayes~\cite{zhang2002recommender}, nearest neighbor methods, and linear classifiers~\cite{adeniyi2016automated} are used to learn a model for the user. DNN-based recommenders have been used to combine CF (through autoencoders) and content-based (training on item attributes) approaches~\cite{wang2014improving,wang2015collaborative,baldi2012autoencoders}. Distributed CF methods have been proposed to solve scalability issues in centralized CF schemes~\cite{ali2004tivo,han2004scalable}. However, all these methods are trained offline, and suffer in performance when the item-set or user-preferences change dynamically. In most social systems of today, it is common to see an item popularity change and evolution in user interests. It therefore becomes essential to learn continuously, requiring online algorithms.

\textbf{Bandit-based recommendation systems:} Many bandit- \\ based approaches for online recommendations exist. Stochastic multi-armed bandits with single and multiple instances of linear bandits were outlined in~\cite{li2010contextual}. Specifically, LinUCB models expected reward through linear regression on context vectors. 
CLUB~\cite{gentile2014online} introduces bandit-based clustering techniques. Dyn\-UCB \cite{nguyen2014dynamic} also propose dynamic clustering using contextual multi-armed bandits. They propose a flat clustering structure, and allow users to switch clusters. Our approach can leverage their clustering technique to allow such user movement between clusters. While CLUB, DCCB, and {\name} only identify user-based clustering, it is possible to simultaneously identify user and item clusters collaboratively such as in~\cite{li2016collaborative}.

Among distributed and decentralized bandits, 
DCCB~\cite{korda2016distributed} proposes  decentralised clustering algorithms to solve linear bandit problems in peer-to-peer networks by sharing information among collaborating workers. However, DCCB suffers scalability issues owing to large transfer volumes.
Tekin et al. propose a decentralized recommendation system~\cite{tekin2014distributed}, where a group of agents can recommend items to users, and privacy and autonomy are important. Agents are prohibited from learning from other agents, and can only use their own experience. This is a fundamentally different constraint than other online systems, such as {\name}, DCCB, and CLUB.

\textbf{RL-based online recommendation algorithms:} Reinforcement learning (RL) techniques are an alternative to bandit-based recommendation systems. Among those, model-based techniques, such as~\cite{modelrl} are unsuited for online recommendations owing to their high complexity~\cite{deeprl2}. 
Deep RL-based techniques are suitable for online recommendations.
Zhao et al.~\cite{zhao2018deep} use a deep Q-network trained with embeddings of users' historical clicked/ordered items (state) and a recommended item (action) as input. Training this state-space generates recommendations. However, such approaches need to evaluate the
Q-values of all the actions under a specific state, which is inefficient for large interaction counts. In general, RL-based approaches involve significant complexity owing to the state space they preserve. For instance, DDR~\cite{deeprl2} stores state transitions in buffers. Such requirements prohibit RL-based techniques from being decentralized, leaving bandit-based approaches alone to be amneable to distribution.

\section{Conclusion}
\label{sec:conclusion}
This paper has presented {\name}, a distributed algorithm and system applicable to contextual bandit algorithms that are prevalent in online recommendation systems where content and user preferences change dynamically. Although DCCB, the state-of-the-art online recommendation system, distributes the computation, it fails to scale effectively. The key reason for this behaviour is the communication bottleneck. {\name} algorithm overcomes this issue by intelligently mixing cluster-based and user-based recommendation schemes. This mixing allows {\name} to generate clusters faster, thereby obtaining better accuracy.

{\name} is implemented using Spark, and efficiently employs distributed primitives in the Spark framework. The paper showed that {\name} can scale to 512 cores with high efficiency. Experimental results over both real-world and synthetic datasets show that {\name} operates on average {\perf} faster than DCCB, and achieves {\accr} better normalized prediction performance. These results indicate that {\name} can be readily deployed in practice.

The presence of large matrix-vector products make this approach highly suitable to be executed on GPUs. We leave GPU porting of {\name} to the future work.

\bibliographystyle{ACM-Reference-Format}
\bibliography{main}

\end{document}